\documentclass{emulateapj}
\usepackage{multirow}
\usepackage{natbib}

\usepackage[3D]{movie15}
\usepackage{hyperref}

\begin{document}

\title{Modeling AGN Feedback in Cool-Core Clusters: The Formation of Cold Clumps}
\author{Yuan Li and Greg L. Bryan}
\affil{Department of Astronomy, Columbia University, Pupin Physics Laboratories, New York, NY 10027}

\begin{abstract}
We perform high-resolution (15-30 pc) adaptive mesh simulations to study the impact of momentum-driven AGN feedback in cool-core clusters, focusing in this paper on the formation of cold clumps.  The feedback is jet-driven with an energy determined by the amount of cold gas within 500 pc of the SMBH.  When the intra-cluster medium (ICM) in the core of the cluster becomes marginally stable to radiative cooling, with the thermal instability to the free-fall timescale ratio $t_{TI}/t_{ff} < 3-10$, cold clumps of gas start to form along the propagation direction of the AGN jets. By tracing the particles in the simulations, we find that these cold clumps originate from low entropy (but still hot) gas that is accelerated by the jet to outward radial velocities of a few hundred km/s.  This gas is out of hydrostatic equilibrium and so can cool.  The clumps then grow larger as they decelerate and fall towards the center of the cluster, eventually being accreted onto the super-massive black hole. The general morphology, spatial distribution and estimated $H{\alpha}$ morphology of the clumps are in reasonable agreement with observations, although we do not fully replicate the filamentary morphology of the clumps seen in the observations, probably due to missing physics.

\end{abstract}

\keywords{}

\section{Introduction}

The intracluster medium (ICM) in the core of many galaxy clusters has low temperature and high density, with a cooling time much shorter than the Hubble time. In a steady state, a cooling flow of 100s M$_{\odot}$/yr is expected (see review by \citet{Fabian1994}). However, Chandra and XMM-Newton have observed a dearth of gas below 2-3 keV \citep[e.g.,][]{Peterson2003, Sanders2008}, suggesting the absence of a classic cooling flow in these cool-core clusters. Various heating mechanisms have been proposed to suppress cooling \citep[e.g.,][]{Soker2001, Narayan2003, Domainko2004, Cattaneo2007, Guo2008, ZuHone2010, Falceta2010}, among which AGN is the most promising due to its self-regulating nature and strong observational support. Numerous X-ray and radio observations have linked AGN activity with cooling \citep[e.g.][]{Dunn2006, Birzan2012}. In addition, it has been shown that the energy in the AGN inflated bubbles is sufficient to offset cooling \citep{Birzan2004}.

Despite the absence of a classic cooling flow, many cool-core clusters are observed to possess cold molecular gas in the center, indicating the existence of a reduced cooling flow at a typical rate of $\sim 10$ M$_{\odot}$/yr, about 10\% of the classic cooling flow rate. This cold gas is observed in the optical (mainly with $H{\alpha}$ emission) \citep{Hu1985, Crawford1999}, infrared ($H_2$ and other line emission) \citep{Donahue2000, Mittal2012}, submillimeter (CO) \citep[e.g.][]{Edge2001} and UV (OVI emission) \citep{Bregman2001, Bregman2006}. A series of surveys by \citet{McDonald10, McDonald11, McDonald12} have found that many cool-core clusters harbor cold gas, some of which only show emission in the nuclei, and some have filamentary structures that extend up to tens of kpc from the center \citep{McDonald10, McDonald11, McDonald12}. Individual filaments observed in nearby cool-core clusters are often long, thin, and clumpy, with a typical width of 100-500 pc and a length of up to a few tens of kpc \citep{Conselice2001}. The kinematics of the extended filaments are usually complex, showing both inflow and outflow, whereas the nuclear emission usually has signatures of rotation \citep{Conselice2001, Salome2006, McDonald12}. The filaments are also found to co-reside with soft X-ray features, UV emissions and dust lanes \citep{Sarazin1992, Fabian2003, Sparks2004, Crawford2005, Fabian2006}.

The correlation between the presence of AGN activity and cold gas \citep{Edwards2007, Cavagnolo2008, Gaspari2012} implies that they are associated; however, despite the rich observations, the precise origin of the cold gas is still unclear. Two popular ideas that have been proposed are: (1) radio bubbles blown by the AGNs rise buoyantly and dredge up cold gas from the bottom of the potential \citep{Churazov2001, Revaz2008, Werner2010}; (2) the gas cools directly out of the cooling flow due to thermal instabilities \citep{Cowie1980, McCourt12, Sharma12, Gaspari2012}. However, it is unclear how bubbles can lift enough cold gas while balancing cooling globally in the long term and it is also unlikely that buoyant bubbles can dredge up the coolest molecular gas from the center of the cluster. \citet{Balbus88} and \citet{Balbus89} have found the ICM to be (linearly) thermally stable, consistent with what we found numerically in \citet{P1} (hereafter Paper I). Both \citet{Sharma12} and \citet{Gaspari2012} introduce external perturbations throughout the cluster to help with the filament formation. 

In this work, we carry out three-dimensional high resolution simulations with Adaptive Mesh Refinement (AMR) code Enzo to study the multi-phase gas in cool-core clusters in the presence of momentum-driven AGN feedback. The key questions we try to address here are: (1) what makes some of the gas cool out of the hot ICM? (2) what are the necessary conditions for this to happen? (3) what is the relationship between the cold gas formation and AGN feedback? (4) how are they related to the cooling of the ICM? For simplicity, we do not include thermal conduction, magnetic fields or star formation. In a companion paper (Paper III), we examine how the AGN feedback can stabilize cooling in our models.

We discuss our methodology in Section~\ref{sec:method} and present our results in Section~\ref{sec:results} including the cold clump morphology, their formation mechanism and internal structure, and how they compare with the observations. In Section~\ref{sec:discussion} we discuss other aspects of the simulations, and in Section~\ref{sec:conclusion} we summarize our results. Throughout this paper, we refer to the diffuse ICM (gas above 1 keV) as being ``hot'' and the gas at temperatures of $\sim 10^4$ K or below as ``cold''.

\section{Methodology}
\label{sec:method}
The simulations are performed using the three-dimensional AMR code Enzo, a parallel, Eulerian hydrodynamics scheme. More details about Enzo can be found in \citet{Enzo}.

As in Paper I, we include radiative cooling in all simulations with the cooling function calculated using Table 4 in \citet{CoolingFunction} assuming half solar metallicity \citep{Metallicity} and equilibrium cooling. The cooling is truncated at $T_{\rm floor} \sim 10^4 K$.

Our standard run in this paper has the number of root grids $N_{\rm root} = 256$, with a box size of $L = 16$ Mpc, and a maximum refinement level of $l_{\rm max} = 12$ during the early stage of the evolution. The physical size of the smallest grid cell in this simulation is $\Delta x_{\rm min} = L / (N_{\rm root} 2^{l_{\rm max}}) \approx 15$ pc. When a large amount of dense clumps have formed at about $t= 380$ Myr, we decrease $l_{\rm \rm max}$ to 11 to save computational time.

\subsection{Initial Conditions}\label{sec:methodology_initial}
The initial conditions of our simulations are very similar to Paper I: we build our idealized galaxy cluster based on the observations of the nearby cool-core Perseus Cluster, assuming spherical symmetry. The initial gas temperature profile follows the analytic fits to the observations of the Perseus Cluster \citep{Churazov} at $r < 300$ kpc:
\begin{equation}
T = 7 \; \frac{1 + (r/71)^3}{2.3 + (r/71)^3} \; \rm{keV}\;,
\end{equation}
where $r$ is the distance to the center of the cluster in kpc. At $r > 300$ kpc, no observed azimuthally averaged temperature profile is available for Perseus. Thus, for the initial gas at $r > 300$ kpc, we adopt the universal temperature profile found by \citet{Universal_T}, normalizing it to match the observations at $r = 300$ kpc:
\begin{equation}
T = 9.18 \times (1+(\frac{3\;r}{2\;r_{\rm vir}})^{-1.6})\; \rm{keV}\;,
\end{equation}
where $r_{\rm vir}$ is the virial radius of the cluster. The gravitational potential in our simulations is static and has three components: the NFW halo, the stellar mass of the brightest cluster galaxy (BCG) and the super-massive black hole (SMBH) in its center. The SMBH is treated as a point mass of $M_{\rm SMBH} = 3.4 \times 10^8$ M$_{\odot}$ at the very center of the simulation domain \citep{BHmass}. The stellar mass profile of the BCG is:
\begin{equation}
M_*(r) = 
\frac{r^2}{G} \left[ \left( {r^{0.5975} \over 3.206 \times 10^{-7}}\right)^s
+ \left( {r^{1.849} \over 1.861\times 10^{-6}}\right)^s 
\right]^{-1/s}
\end{equation}
in cgs units with $s = 0.9$ and $r$ in kpc \citep{Mathews}. We also adopt the NFW parameters $M_{\rm vir} = 8.5 \times 10^{14}$ M$_{\odot}$, $r_{\rm vir} = 2.440$ Mpc and concentration $c=6.81$ from \citet{Mathews}, where they fit the NFW profile based on the observed temperature and density profiles of the Perseus Cluster assuming hydrostatic equilibrium. Note that, strictly speaking, the NFW halo here is not just the dark matter halo, but also includes the gravity of the intra-cluster gas. We neglect self gravity of the gas in our simulations as it does not contribute significantly.  As we have shown in Fig. 1 of Paper I, the gravity of the gas is not dominant at any radii (at least at early times).

Given the initial gas temperature profile and the gravitational potential, we compute the initial gas density and pressure assuming hydrostatic equilibrium and the ideal gas law with an adiabatic index $\gamma = 5/3$. Since we are focusing on the effects of AGN feedback, unlike Paper I, we do not introduce any initial perturbation or rotation of the gas. We also turn off the Hubble expansion to avoid its possible effect on the long term evolution of the cluster.

\subsection{Jet Modeling}\label{sec:methodology_jet} 
The SMBH feedback loop is triggered by the accretion of gas onto the black hole. In our simulations, we only consider the accretion of cold gas. The jet power is calculated based on the estimated accretion rate. Ideally, the accretion rate should be computed near the event horizon of the black hole, but since our simulations do not resolve the size of the event horizon or the actual accretion disk, and we do not have the physics for the accretion disk, we estimate the accretion rate based on some assumptions. We first sum up the mass of all the cells that are ``cold'' ($T< T_{\rm cold}$) within a box of $1$ kpc$^3$ in the center of the simulation domain, which contains the sphere of influence of the SMBH, and define this as $M_{\rm cold}$. We assume that this cold gas will be accreted within a typical accretion time $\tau$ to obtain the estimated accretion rate $\dot{M} =  M_{\rm cold}/\tau$. We then remove the cold gas that is accreted to the SMBH by decreasing the mass of the cold cells in proportion to its mass normalized such that the total mass removed inside the accretion box is equal to $\dot{M} \Delta t$ at each time step $\Delta t$. We use $T_{\rm cold} = 3.0 \times 10^4$ K in our simulations, a few times the temperature of the cooling floor ($T_{\rm floor} \sim 10^4$ K). The exact value of $T_{\rm cold}$ is not important because gas spends very little time at temperatures between $10^7$ and $10^4$ K once it starts runaway cooling.  We adopt an accretion time $\tau = 5 \rm$ Myr in these simulations, but again, the choice of the exact value of $\tau$ also has little effect on the overall results of the simulations since $\tau$ is only the timescale over which the accretion rate is averaged.

Based on the estimated accretion rate $ \dot{M}$, we can compute the total power in the jets as 
\begin{equation}\label{eq:Edot}
\dot{E} = \epsilon  \dot{M} c^2 ;
\end{equation}
where $\epsilon$ is the effective feedback efficiency of the SMBH. Mass conservation relates $\dot{M}_{\rm SMBH}$, the rate at which materials fall into the black hole or the black hole growth rate, and the accretion rate $\rm \dot{M}$ by $\dot{M}_{\rm SMBH} = \dot{M}-\dot{M}_{\rm jet} = \epsilon_1 \dot{M}$ , with $ \dot{M}_{\rm jet}$ being the outflow rate or the mass loss rate in the jets. The power of the black hole is then $ \dot{E} = \epsilon_2 \dot{M}_{\rm SMBH} c^2 = \epsilon_1 \epsilon_2 \dot{M} c^2$. So our effective efficiency $\epsilon = \epsilon_1 \epsilon_2$ is actually a combination of the black hole growth efficiency $\epsilon_1$ and the black hole accretion efficiency $\epsilon_2$. A typical value for $\epsilon$ is $10^{-4}-10^{-2}$ \citep[e.g.,][]{Gaspari2011}. We choose a moderate value of $\epsilon=0.001$ in the standard run discussed in this paper -- the process of clump formation is not sensitive to the precise value chosen.  The black hole growth rate is expected to be small in radio mode AGN. Therefore, we approximate $\dot{M}_{\rm jet} \approx \dot{M}$.

We launch a pair of symmetric jets in two circular planes of radius $2 r_{jet}$ at a distance of $h_{jet}$ parallel to the xy plane. During each time step $\Delta t$, mass $\Delta m$ is added to the cells within the jet launching planes so that, for the cell at distance r from the z-axis, $\Delta m \propto e^{-r^2/2r^2_{\rm jet}}$ and the total $\Delta m$ in the two planes adds up to the total mass accreted: $\int \Delta m = \dot{M} \Delta t$. 

The total amount of kinetic energy in the jets $\dot{E}_{\rm kinetic}=\dot{E}-\dot{E}_{\rm thermal}=f\dot{E}$. Since we do not resolve the bottom of the jets and it is not clear what fraction of the jet energy is in kinetic form, we have experimented with pure kinetic feedback ($f=1$) and half-thermalized feedback ($f=0.5$). In the primary simulations we discuss in this paper, we have used $f=0.5$. The kinetic energy of the jets is related to the jet velocity $v_{\rm jet}$ by $\dot{E}_{\rm kinetic}= \frac{1}{2}\dot{M}v^2_{\rm jet}$. Therefore, $v_{\rm jet}=\sqrt{2f\epsilon}c \approx 10^4$ km s$^{-1}$ for $\epsilon=0.001$ and $f=0.5$.

In each cell of the jet launching planes, the added mass $\Delta m$ has a momentum of $v_{\rm jet} \Delta m$, and the resulting velocity of the cell is computed according to the conservation of momentum; the thermal energy added to the cell is $\frac{1-f}{f}\times \frac{1}{2} \Delta m v^2_{\rm jet}$. 

We add a nested static refine region around the jet launching region to make sure that it is always refined to the highest refinement level of the simulations. In the main simulation discussed in this paper, the radius of the jet launching plane is $2r_{\rm jet}=6 \Delta x_{\rm min}\approx 90$ pc, and the distance of the plane to the SMBH is $h_{\rm jet}=10 \Delta x_{\rm min} \approx 150$ pc.

Simple straight jets in high resolution hydro simulations have been found to create low density channels, causing most of the energy to flow to large radii and therefore fail to balance cooling \citep[e.g.,][]{Reynolds06}. Different generations of bubbles are observed to have different orientations in nearby cool-core clusters, possibly due to re-orientation of the jets \citep[e.g.,][]{Babul2013}. Therefore, in our simulations, we force the jets to precess along the z-axis with a precession period of $\tau_p$, adopting an angle $\theta$ to the z-axis. In the simulations in this paper, $\tau_p = 5$ Myr and $\theta=0.15$. Varying $\tau_p$ or $\theta$ does not significantly change the properties of the cool clumps.

\subsection{Tracer Particles}\label{sec:methodology_tp}
In order to follow the evolution of fluid elements, especially the gas that cools into clumps, we inject tracer particles to the simulations. These tracer particles are Lagrangian, massless particles that passively advect with the flow in the cells, taking the velocity of the cells they reside in. They do not affect the properties (density, temperature, velocity, etc) of the gas in the cell but simply record them at each time step. 

To decide when and where to inject tracer particles, we first run a simulation without tracer particles, finding that a significant number of cold clumps start to form within $r < 10$ kpc at about $t_1=330$. We then re-run the simulation, restarting at $t_1$, but now adding tracer particles and using more frequent data outputs. Given the approximate symmetry of our simulations, we only inject tracer particles into the first octant of the central $10$ kpc sphere. The total number of cells in this region (and thus the total number of tracer particles added) is about 1.5 million.

\section{Results}
\label{sec:results}

We now describe our results, focusing on the formation of cold clumps, which we find to be a robust and generic feature of our simulations. In Section~\ref{sec:morphology}, we present the general properties of these clumps in our standard simulation including their morphology and spatial distribution. In Section~\ref{sec:formation}, we discuss the formation of the clumps. The internal structure of the clumps is shown in Section~\ref{sec:structure}, and in Section~\ref{sec:observations}, we compare our results with the observations.

\subsection{Clump Morphology}\label{sec:morphology}

During the first 150 Myr, the density in the center of the cluster increases as the temperature decreases through radiative cooling. No local instabilities develop at this stage. More details of the early evolution of the pure cooling flow are discussed in Paper I. The SMBH feedback is turned on when runaway cooling starts to happen in the very center of the cluster, about 150 Myr after we start our simulation from the initial configuration. At first, there is little cold gas, so the jet is weak, allowing the average temperature of the ICM to continue decreasing in the cluster core; at this point, runaway cooling (down to $T_{floor}$) only occurs in the immediate vicinity of the SMBH, with no local thermal instabilities developing at larger radii. Due to the increased amount of cold gas near the SMBH, the jet power steadily rises from a few times $10^{42}$ erg s$^{-1}$ to a few times $10^{44}$ ergs s$^{-1}$ within about two hundred Myr. At this point, cold clumps begin to form outside of the very center of the cluster, at radii of 5-10 kpc, about 330 Myr after we start the simulation. The spatial distribution of the clumps extends to larger radii (up to $\sim 20$ kpc) as both the total number and the size of individual clumps grow with time.

\begin{figure*}
\begin{center}
\includegraphics[scale=.22,trim=0cm 0cm 4.2cm 0cm, clip=true]{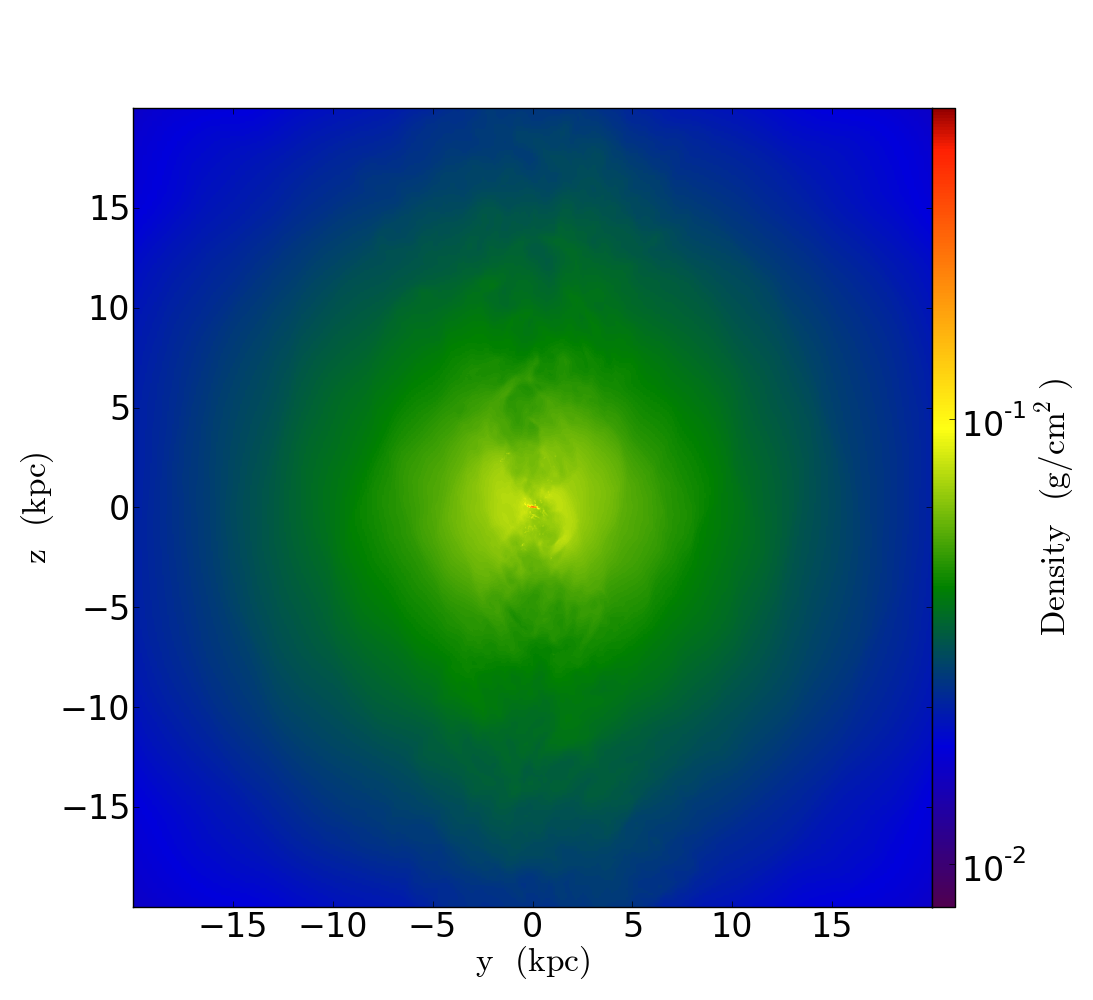}
\includegraphics[scale=.22,trim=3.7cm 0cm 4.2cm 0cm, clip=true]{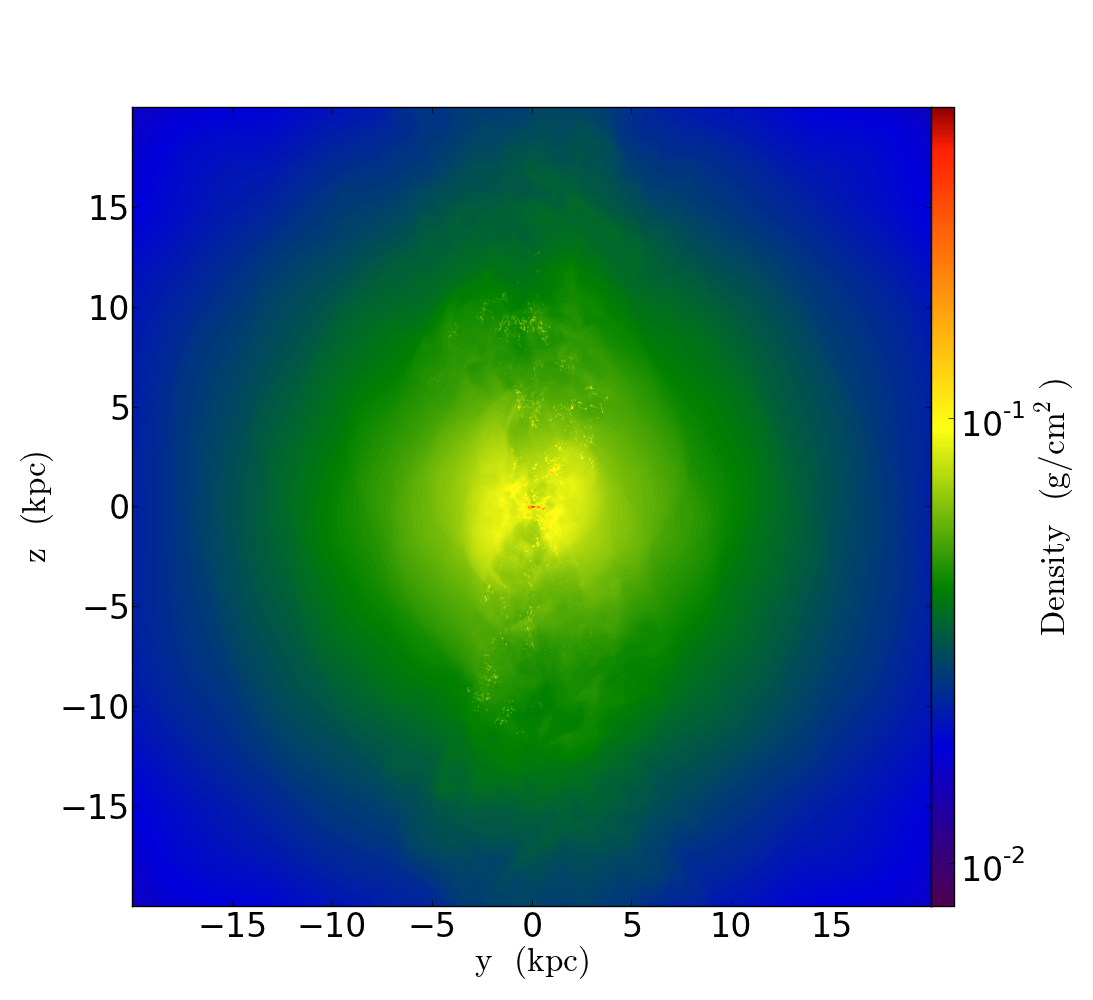} 
\includegraphics[scale=.22,trim=3.7cm 0cm 0cm 0cm, clip=true]{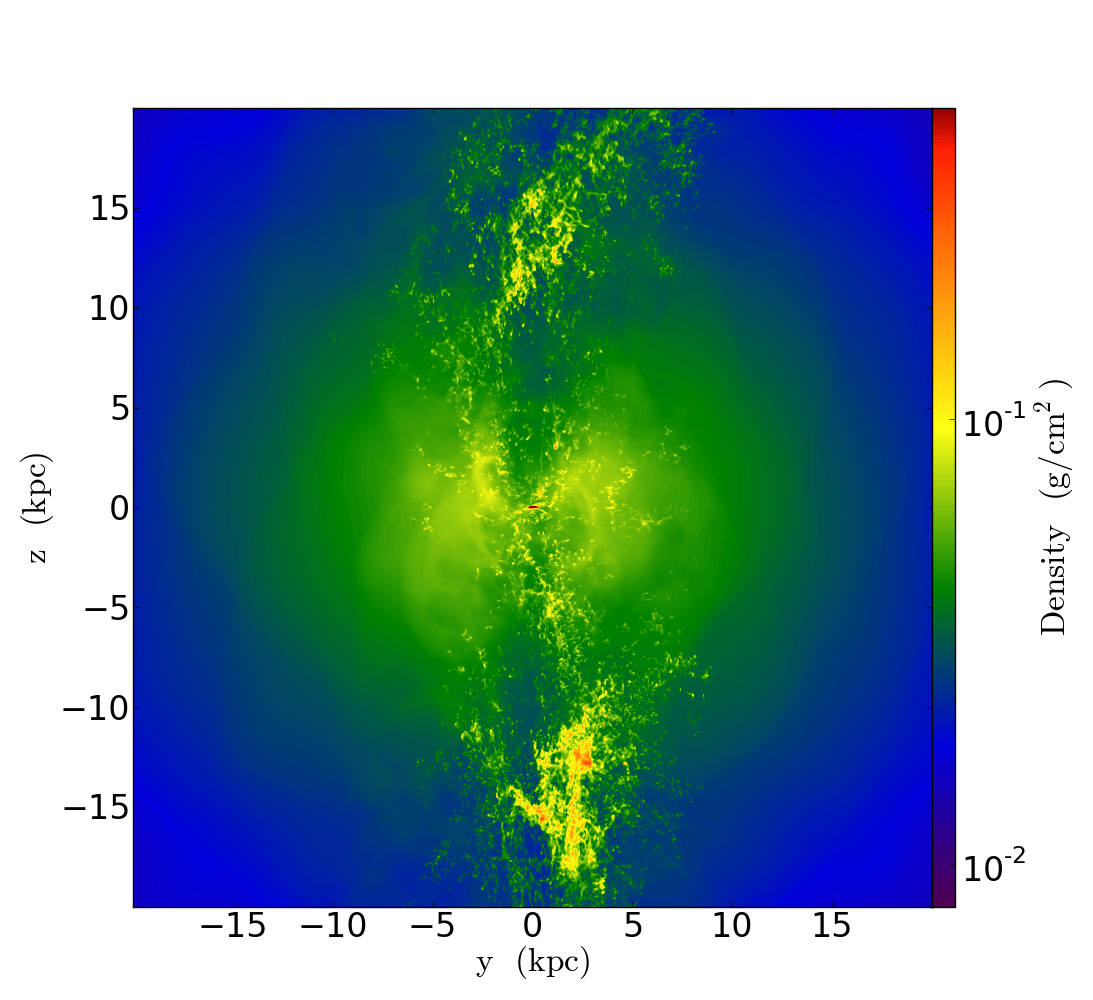}
\caption{The gas density projected along the x-axis at $t_1 = 330$ Myr (left), $t_2 = 356$ Myr (middle), and $t_3 = 410$ Myr (right). The width and depth of the projection are both 40 kpc.
\label{fig:projection_density}}
\end{center}
\end{figure*}

\begin{figure*}
\begin{center}
\includegraphics[scale=.3,trim=0cm 2.9cm -2cm 0cm, clip=true]{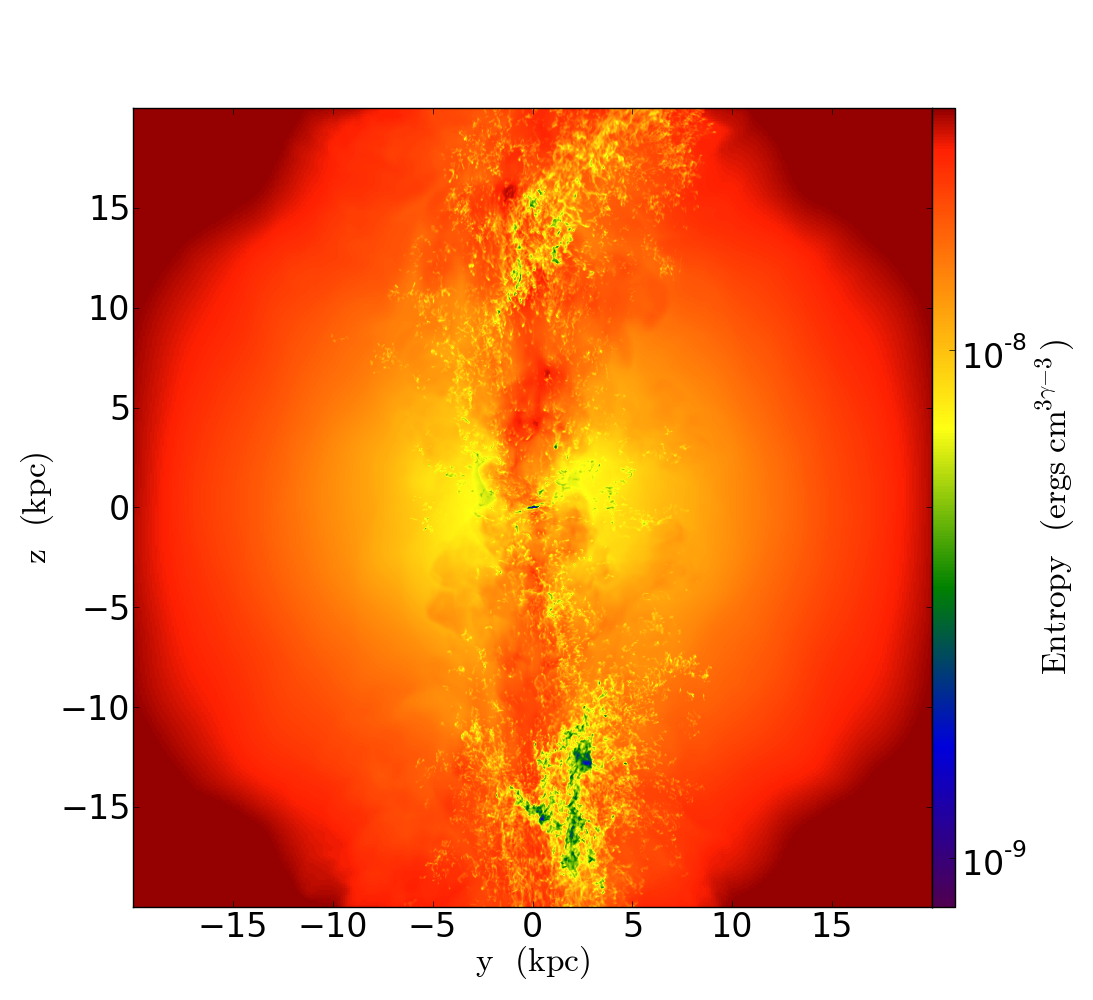}  
\includegraphics[scale=.3,trim=3.45cm 2.9cm 0cm 0cm, clip=true]{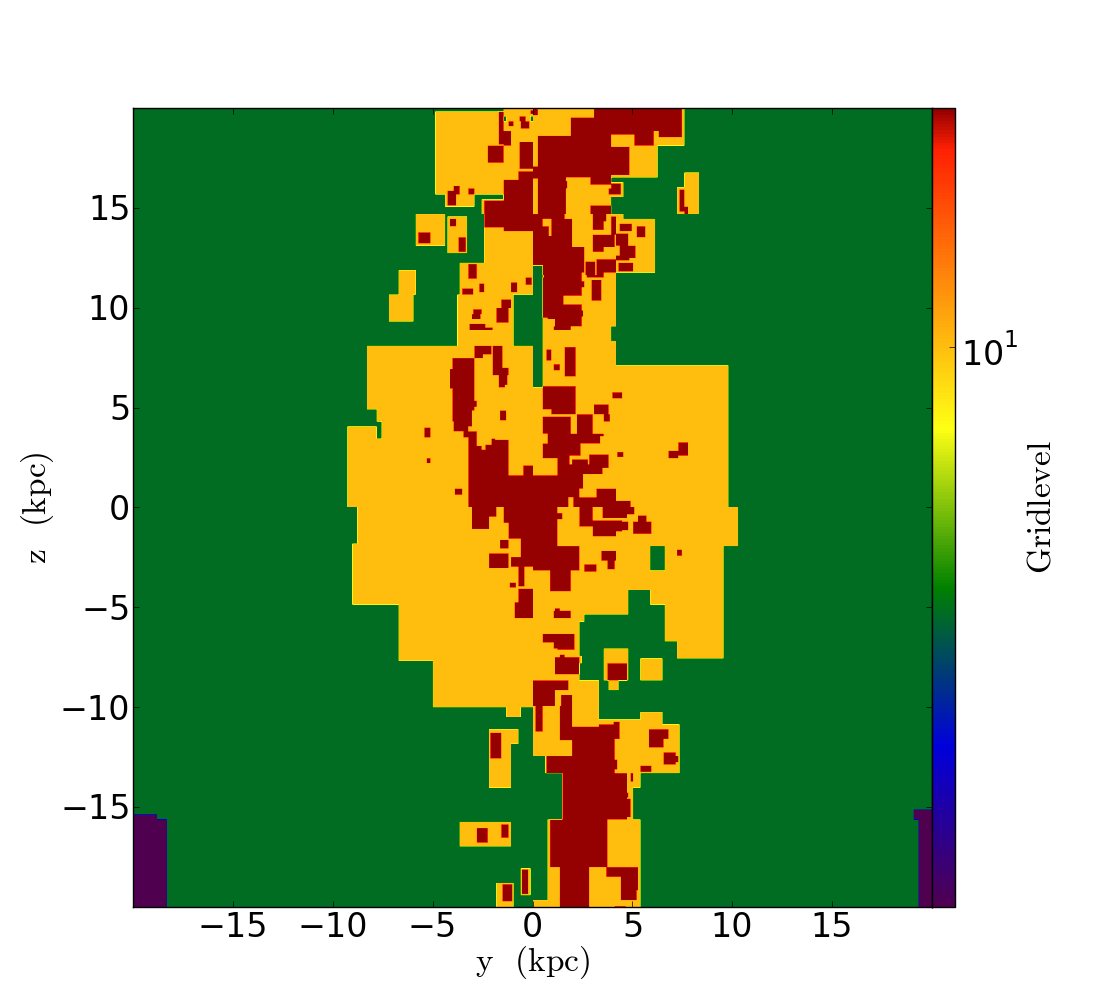}
\includegraphics[scale=.3,trim=0cm 2.9cm -2cm 2.45cm, clip=true]{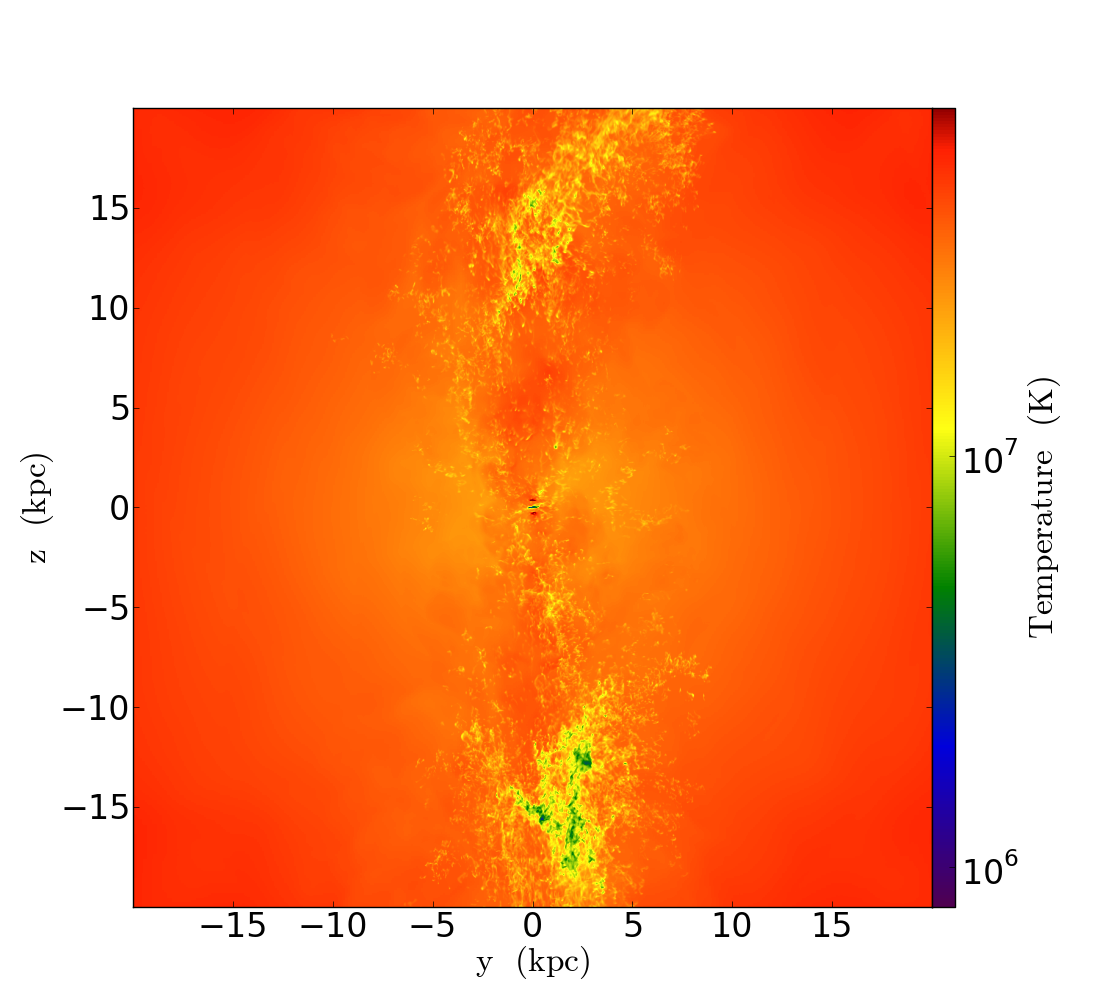}
\includegraphics[scale=.3,trim=3.45cm 2.9cm 0cm 2.45cm, clip=true]{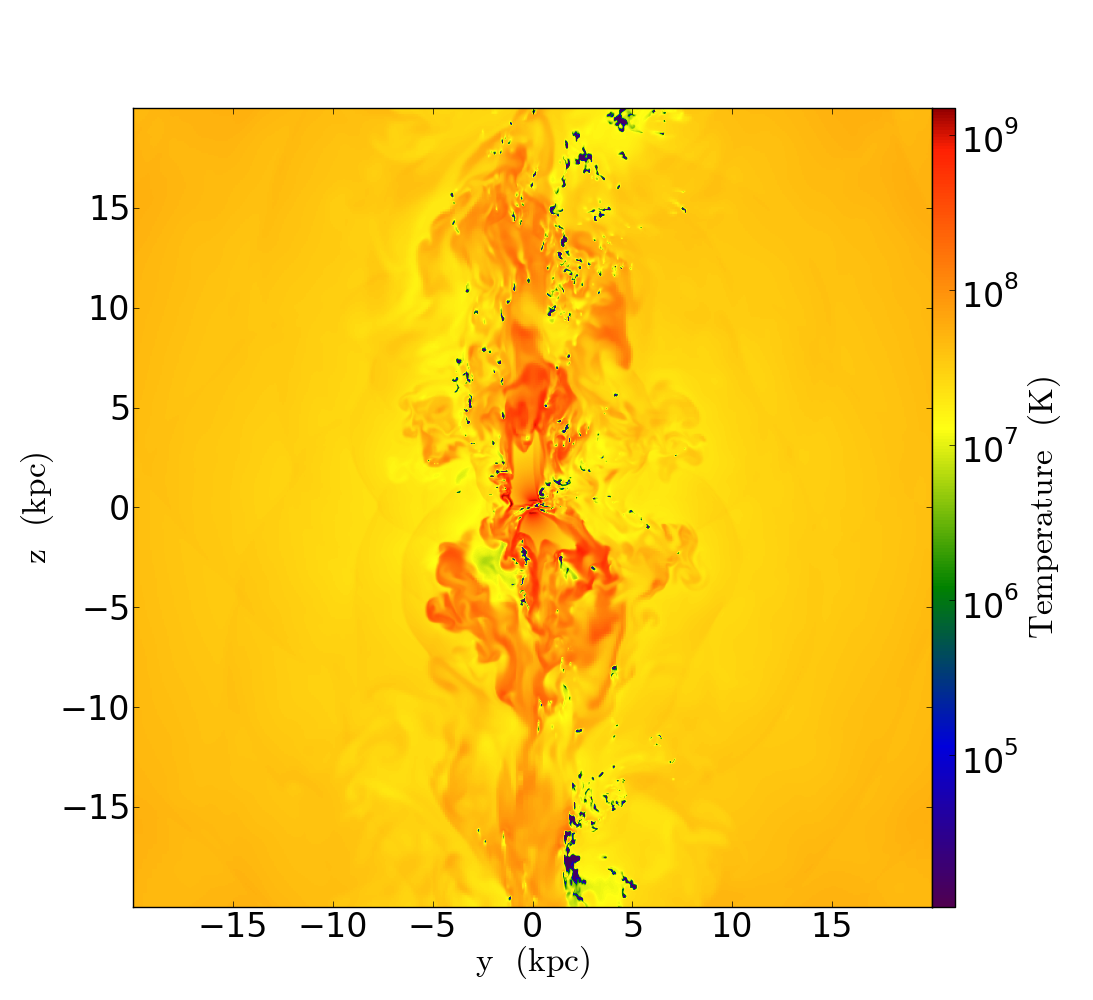}
\includegraphics[scale=.3,trim=0cm 0cm -2cm 2.45cm, clip=true]{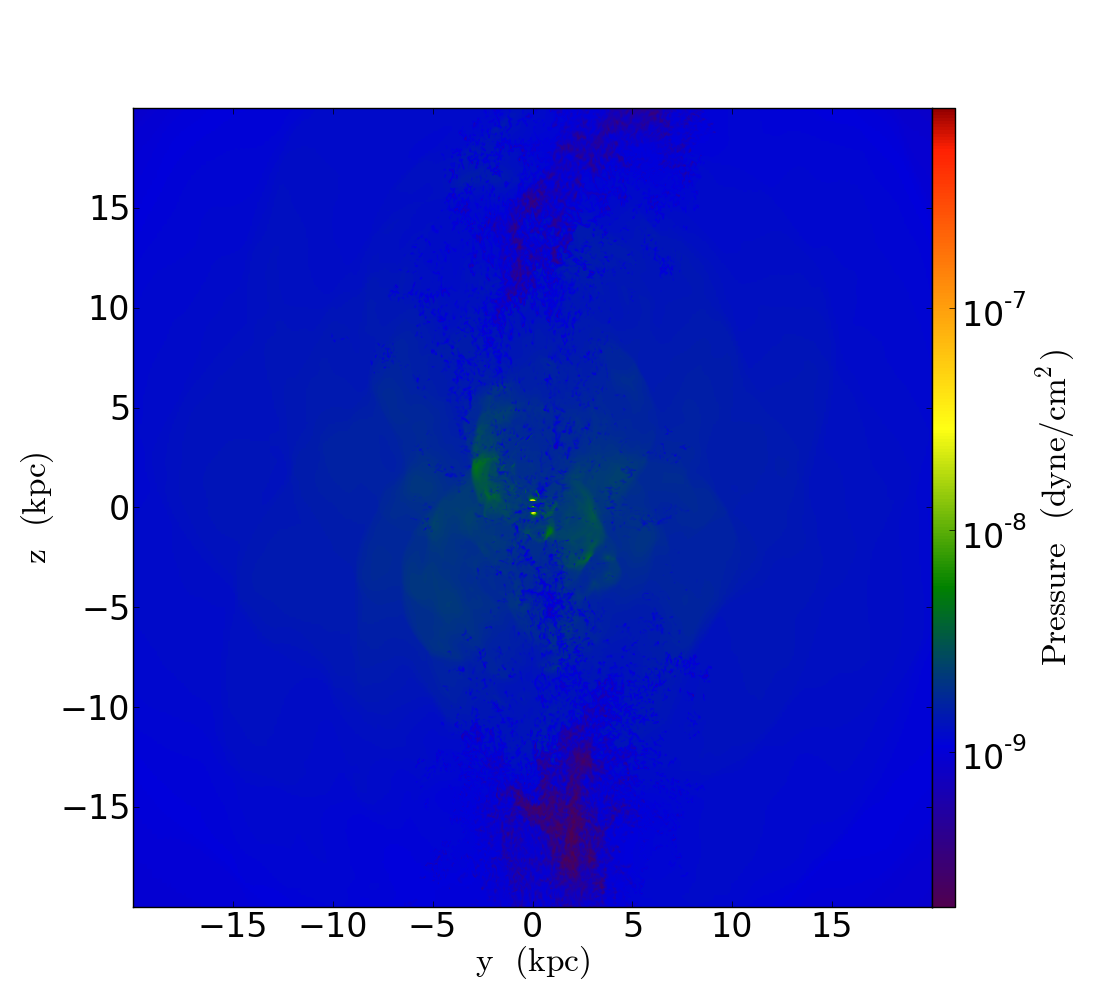}
\includegraphics[scale=.3,trim=3.45cm 0cm 0cm 2.45cm, clip=true]{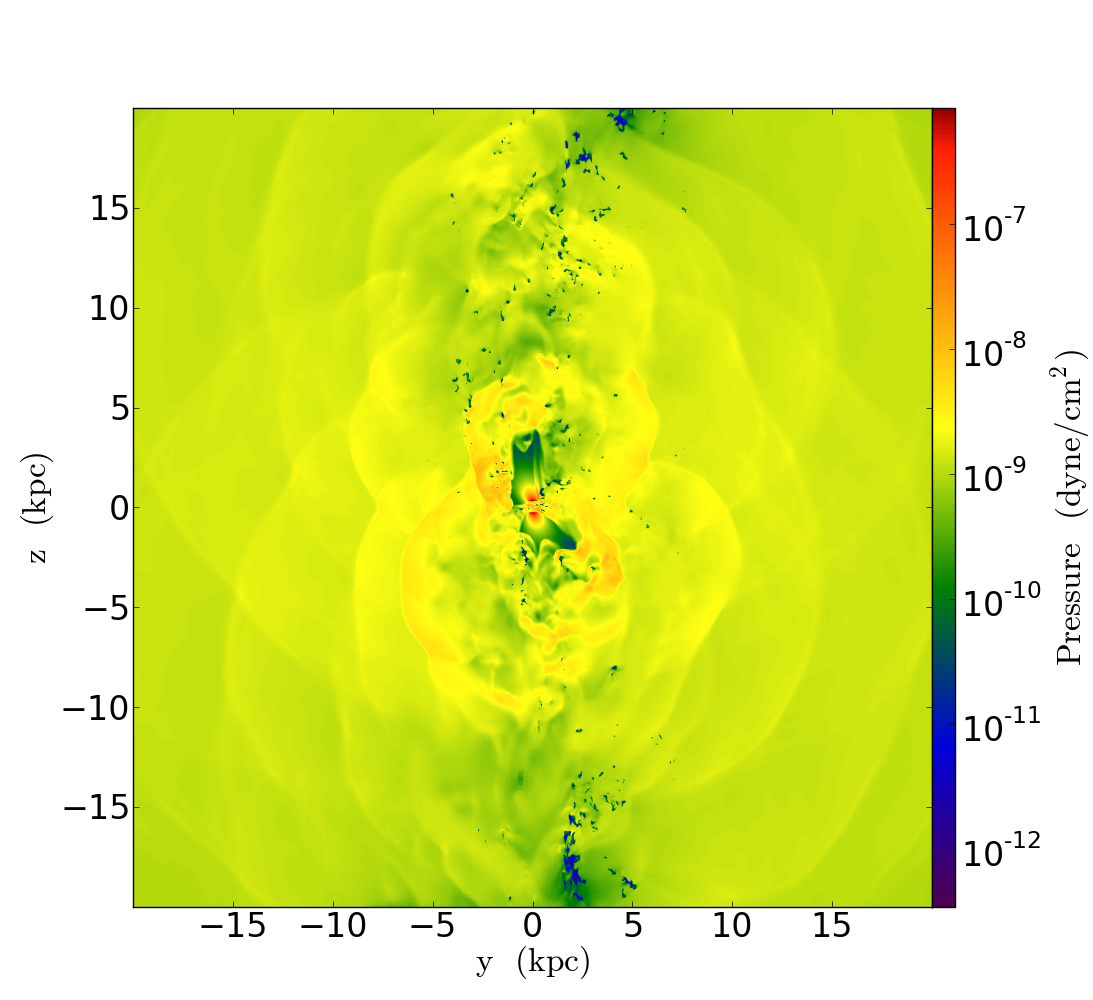}
  
\caption{Left column from top to bottom: mass weighted gas entropy, temperature and pressure projected along x-axis at $t_3$. Right column from top to bottom: the refinement level of the simulation, gas temperature and pressure of a slice of gas through the center of the cluster in the y-z plane. The size of the projection box is 40 kpc, the same as Figure~\ref{fig:projection_density}. 
\label{fig:projection}}
\end{center}
\end{figure*}

We show in Figure~\ref{fig:projection_density} the projected gas density from three different outputs (at $t=$ 330, 356 and 410 Myr, hereafter referred to as $t_1$, $t_2$ and $t_3$) in the central 40 kpc region. The first output ($t_1$) is when clumps just begin to form and is also the point at which we inject tracer particles as discussed in Section~\ref{sec:methodology_tp}.  The second output ($t_2$) is 26 Myr later, when a substantial number of clumps have formed (and is when we stop our tracer particle run).  Finally, the third output ($t_3$) is 80 Myr later, by which point a large amount of clumps are present.  

Figure~\ref{fig:projection} shows the x-projections of mass-weighted gas temperature, pressure and entropy (defined as $k_{B}T/n^{\gamma-1}$ where $k_B$ is the Boltzman constant and $n$ is the particle number density), along with the refinement level, temperature and pressure in a slice of gas in the y-z plane through the center of the cluster at $t_3$ in the central 40 kpc region. The clumps have drastically higher density, lower temperature, entropy and pressure than the surrounding ICM, and are refined to the highest level of resolution due to their high density. 

Figure~\ref{fig:3d} is an interactive 3D model showing the complex structure and morphology of the clumps ($\rho \geq 10^{-23}$ g cm$^{-3}$) at $t=$ 360 Myr which roughly corresponds to $t_2$ (the middle panel of Figure~\ref{fig:projection_density}). The colors represent the surface temperature of the clumps. Many clumps are elongated in the radial direction, with a hotter head (the end closer to the center of the cluster) being shock-heated from the jets and a cooler tail. The internal structure of individual clumps is discussed in Section~\ref{sec:structure}. 

It is already conspicuous from Figure~\ref{fig:projection_density}, \ref{fig:projection} and \ref{fig:3d} that most of the cold clumps are located along the z-axis, the jet propagation direction. We further show this in Figure~\ref{fig:theta} where we plot the distribution of the $(r, \theta)$ of the cold clumps by selecting the cells with temperatures of $T<3\times 10^4$ K. The majority of the cold clumps are located within $30^\circ$ from the z-axis, at a distance between 10 and 20 kpc from the center of the cluster.

\begin{figure*}
\begin{center}
\includemovie[
    poster,
    toolbar, 
    label=out2.u3d,
    text=(out2.u3d),
    3Daac=60.000000, 3Droll=0.000000, 3Dc2c=-99.500000 -1641.500000 37.500000, 3Droo=1644.940308, 3Dcoo=-99.500000 21.500000 -37.500000,
    3Dlights=CAD,
]{\linewidth}{\linewidth}{out2.u3d}
\caption{An interactive 3D representation of the clumps ($\rho \geq 10^{-23} \; g/cm^3$) at $t_2$, color coded based on the surface temperature.\label{fig:3d}}
\end{center}
\end{figure*}

\begin{figure}
\begin{center}
\includegraphics[scale=.35,trim=1cm 0cm 0cm 0cm, clip=true]{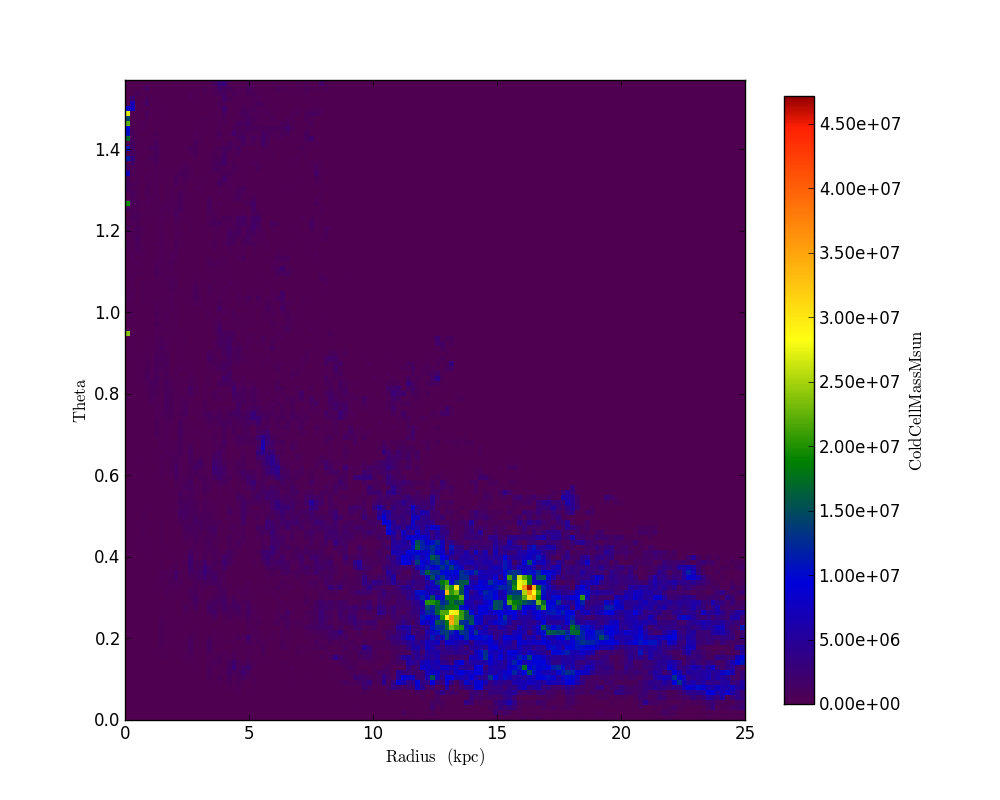}
\caption{The $(r, \theta)$ distribution of the cold clumps at $t_3$, the same time as Figure~\ref{fig:projection}.\label{fig:theta}}
\end{center}
\end{figure}

\subsection{Clump Formation}
\label{sec:formation}

\begin{figure}
\begin{center}
\includegraphics[scale=.40,trim=1.0cm 0cm 0cm 0cm, clip=true]{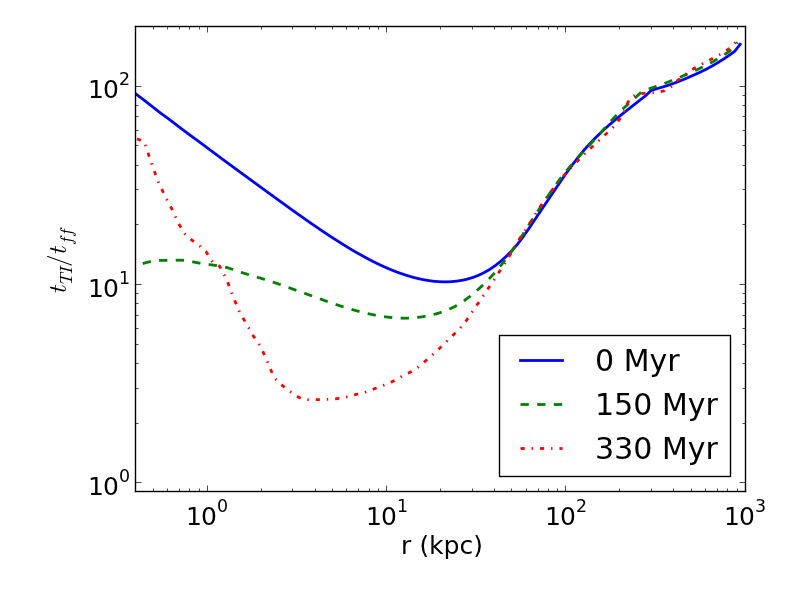}\\
\includegraphics[scale=.40]{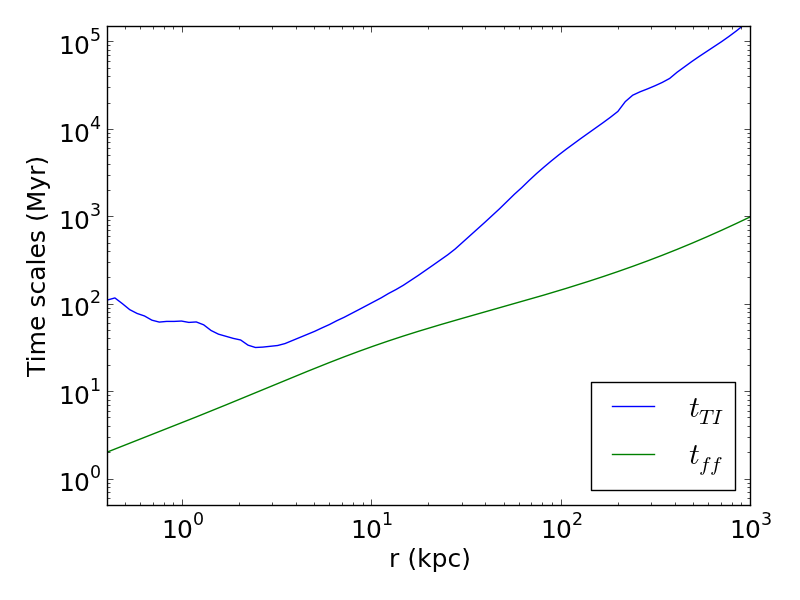}
\caption{Left: the ratio of the thermal instability timescale to the free-fall timescale of the ICM $t_{TI}/t_{ff}$ at different stages. Right: the thermal instability and the free-fall timescales at $t_1$, which corresponds to the dash-dotted red line on the left panel. \label{fig:ratio}}
\end{center}
\end{figure}

We now explore how the gas clumps form.  \citet{McCourt12} found that local thermal instabilities can grow and produce cold clumps in gravitationally stratified plasma in plane-parallel simulations when $t_{TI}/t_{ff} \lesssim 1$, where
$t_{TI}=\frac{5}{3}\frac{t_{\rm cool}}{2-d \ln \Lambda /d \ln T}$
is the thermal instability timescale and $t_{ff}=\sqrt{2r/g}$ is the free-fall timescale. \citet{Sharma12} found this critical ratio to be $\sim 10$ in a spherical potential. We plot in Figure~\ref{fig:ratio} the evolution of the azimuthally averaged $t_{TI}/t_{ff}$ of the cluster in our simulation. Initially this ratio is above 10 throughout the cluster. As we have already pointed out in Paper I, the local instability does not develop in the pure cooling stage. When the (global) cooling instability begins in the very center ($r\lesssim 100$ pc) of the cluster, and feedback is turned on at $t\sim150$ Myr, this ratio is still above 7 outside of the cooling center, and gas condensation outside of the central few hundred pc is not seen for another 200 Myr despite the perturbation from the jets. Cold clumps finally start to form at about $t_1 = 330$ Myr, which corresponds to the left panel of Figure~\ref{fig:projection_density}, when the lowest $t_{TI}/t_{ff}$ outside the central cooling region first drops below $2-3$, shown as the red dot-dashed line in the left panel of Figure~\ref{fig:ratio}. The right panel of Figure~\ref{fig:ratio} shows the azimuthally averaged $t_{TI}$ and $t_{ff}$ at this time. The minimum of the ratio $t_{TI}/t_{ff}$ as well as $t_{TI}$ itself occurs at radii between $3-8$ kpc, which is roughly where the clumps first form ($r\sim 5-10$ kpc, see Figure~\ref{fig:projection} and Figure~\ref{fig:theta}). This indicates that low $t_{TI}/t_{ff}$ or short $t_{TI}$ is a necessary condition for the development of the local cooling instability. 

However, this is not a sufficient condition for cold clumps to form. As shown in Section~\ref{sec:morphology}, clumps appear almost exclusively along the jet propagation direction even though the ICM in our simulation is roughly spherically symmetric (as seen in Figure~\ref{fig:projection_density} and Figure~\ref{fig:projection}). To further test this spherical symmetry, we have compared the average $t_{TI}/t_{ff}$ within the jet propagation cone (we use a cone of $30^{\circ}$) and that outside the cone, and find no significant difference. In fact, the ratio is even slightly higher within the jet cone due to the shock heating. Therefore, the jet must create another necessary condition for the formation of the clumps.

To find out how the jet triggers clump formation, or in other words, what makes the originally hot gas cool into clumps along the jet propagation direction, we inject tracer particles in the simulation as discussed in Section~\ref{sec:methodology_tp}. We first identify the particles that are cold (with $T<3\times10^4 K$) close to $t_2$, the end of our tracer particle run. Then, among these particles, we select the ones that were once hot (with $T>5\times 10^6$ K) in the past because we only want to analyze the particles that undergo a phase transition (from hot to cold) and we do not want the ones that have always been cold to contaminate our analysis. We also exclude the particles that were originally within 500 pc from the SMBH where runaway cooling has been going on for a while.  The contamination from both is minimal as the amount of cold gas at $t_1$ is much lower than at $t_2$. Information on the physical properties of these particles is collected at the times before and after the phase transition and is compared against the whole sample of all the particles. 

\begin{figure*}
\begin{center}
\includegraphics[scale=.43,trim=1.4cm 0cm 3.5cm 0cm, clip=true]{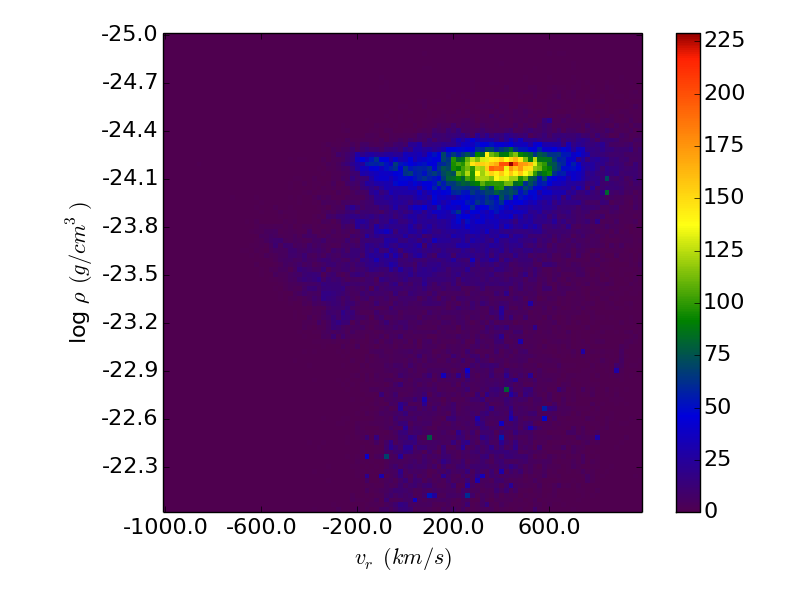}
\includegraphics[scale=.43,trim=2.4cm 0cm 3.5cm 0cm, clip=true]{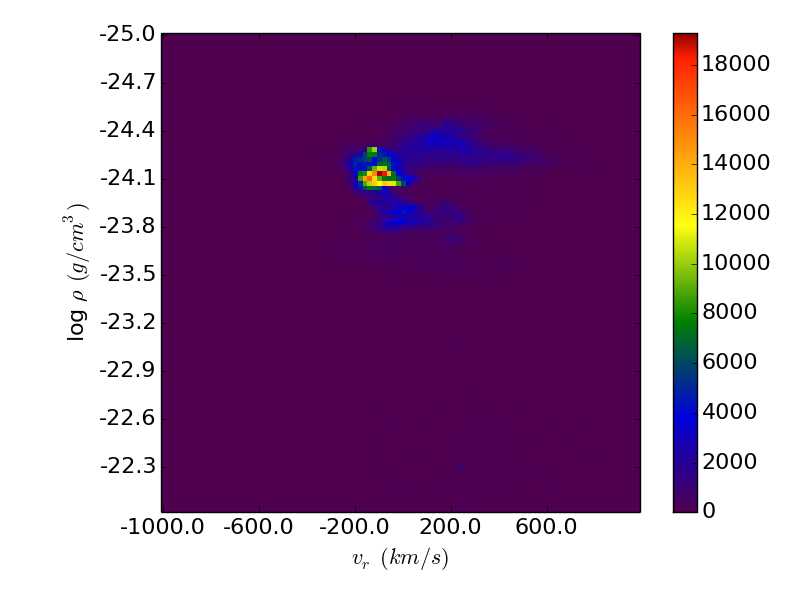}
\caption{The density ($\rho$) - radial velocity ($v_r$) distribution of the particles right before they cool (left) and that of the whole sample at $t_2$ (right).\label{fig:rho_Vr}}
\end{center}
\end{figure*}

\begin{figure}
\begin{center}
\includegraphics[scale=.39]{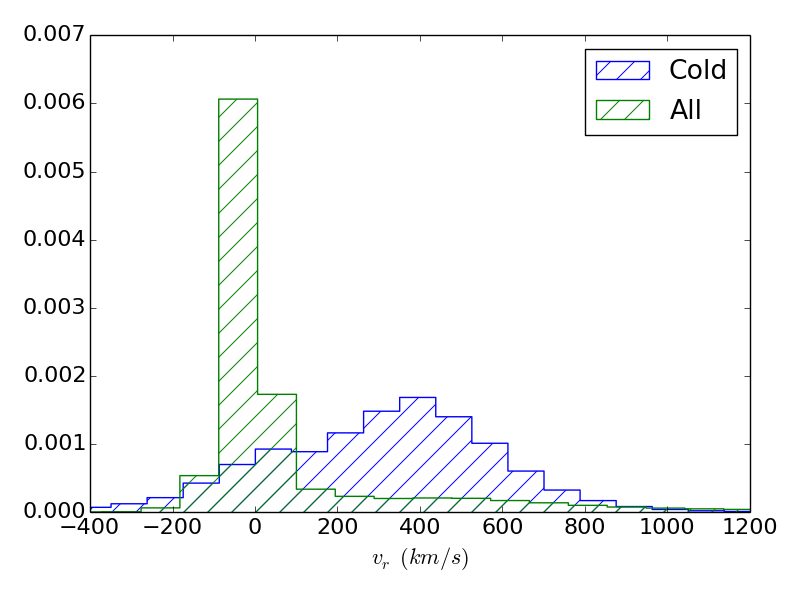}
\caption{The 1D normalized distribution of the radial velocities ($v_r$) of particles in the same two samples as in Figure~\ref{fig:rho_Vr}. The distribution of the particles that are about to cool is shown in blue and the whole sample is shown in green, the former showing a systematically more positive radial velocity and the later.\label{fig:Vr}}
\end{center}
\end{figure}

In Figure~\ref{fig:rho_Vr}, we plot the density ($\rho$) - radial velocity ($v_r$) distribution of each particle immediately before it cools (0.4 Myr before the temperature drops below $T_{\rm cold}$) and compare it with the whole sample of tracer particles at $t = 350$ Myr.  The density distribution of the particles that are just about to cool does not differ significantly from the whole sample. This is also true for the temperature and pressure distributions (not shown). However, the radial velocities of those particles are generally more positive than those of the whole sample. This is shown more clearly in Figure~\ref{fig:Vr} where we plot the 1D normalized distribution of the radial velocities of the two samples.

Linear perturbation analysis has found that the ICM in cool-core clusters is thermally stable \citep[e.g.,][]{Balbus88, Balbus89}, which is consistent with what we have found in our previous simulations focusing on the pure cooling flow, in Paper I. When the thermal instability timescale is shorter than the free-fall time, the gas is stable \citep[e.g.,][]{Joung2012} and a small perturbation to a gas parcel will damp as it oscillates in the direction of gravity. In our simulation, when clumps just start to form, $t_{TI}/t_{ff}$ is above 1 (see Figure~\ref{fig:ratio}), so the gas is linearly stable and small perturbations oscillate but do not grow. That is consistent with what we see outside the jet propagation cone: there is weak turbulent motion, but no clumps are formed. Along the jet propagation direction, however, some of the gas parcels are accelerated by the jet to moderate positive radial velocities -- a few hundred km/s -- which does not allow them to return back to a stable (oscillating) state, and therefore they can cool through radiation and expansion due to the decrease of ambient pressure. 

Another way to understand this process is by examining the right panel of Figure~\ref{fig:ratio}. The minimum of $t_{TI}$ occurs at $r \sim 3$ kpc, and is about 3 times longer than the local $t_{ff}$. In an adiabatic process, $t_{TI} \propto S^{\frac{6}{5}}P^{-\frac{1}{5}} \propto P^{-\frac{1}{5}}$ (for constant entropy $S$). As we have analysed in Paper I, from a few hundred pc to a few tens of kpc, $\rho \propto g \propto r^{-0.75}$ in hydrostatic equilibrium. Therefore, $P \propto r^{-0.5}$ and $t_{TI} \propto r^{0.1}$. When a gas parcel is lifted by the jet from $r\sim 3$ kpc adiabatically, its $t_{TI}$ stays roughly constant while $t_{ff}$ increases quickly with r ($t_{ff} \propto r^{0.875}$), and therefore $t_{TI}$ of the clump moves closer to the local $t_{ff}$ as the gas is raised to larger radii. The actual process of uplifting is not adiabatic, and $t_{TI}$ shortens when radiative cooling is taken into consideration. Given the typical radial velocity of a few hundred km/s and $t_{TI}\sim 30$ Myr, the gas travels $\sim 5-10$ kpc before it cools. This also explains why the clumps are first found at radii slightly larger than where the minimum $t_{TI}/t_{ff}$ ratio is located.

Shock compression can also help with clump formation because the cooling time decreases after weak shocks; however shocks do not appear to be the determining factor as the densities of the particles before they cool do not differ significantly from the whole population, as we have shown in Figure~\ref{fig:rho_Vr}. Also, shock waves propagate in all directions in our simulation (seen in the pressure plot in Figure~\ref{fig:projection}), but clumps only form in the jet direction. This indicates that shock compression alone cannot explain the formation of clumps in our simulations.

Note that the critical $t_{TI}/t_{ff}$ for thermal instability to develop in our simulations is slightly lower than what \citet{Sharma12} found in their cluster simulations ($t_{TI}/t_{ff} \lesssim 10$). This is likely due to the difference in the feedback mechanism in our simulations: \citet{Sharma12} imposes a heating term which balances cooling for each radius, wheras our heating is solely driven by the jets.  At early times in our simulation, the turbulent level is still low and therefore a lower $t_{TI}/t_{ff}$ ratio is required. In fact, after the initial burst of clump formation discussed here, new clumps are still formed (at a much lower rate) despite the increased ratio of $t_{TI}/t_{ff}$. This is because the turbulence is stronger due to the increased jet power.

\subsection{Clump Structure}\label{sec:structure}

To better understand the inner structure of individual clumps we identify all the clumps with the yt \citep{yt} clump finding method, the details of which can be found in \citet{Britton09}. The clumps are identified at all three times from $t_1$ to $t_3$ as regions with density $\rho > 10^{-22}$ g cm$^{-3}$ and temperature $T < T_{\rm cold} = 3 \times 10^4$ K. They are also required to contain at least 6 cells to remove very poorly resolved objects. 

\begin{figure}
\begin{center}
\includegraphics[scale=.39,trim=1.5cm 0cm 0cm 0cm, clip=true]{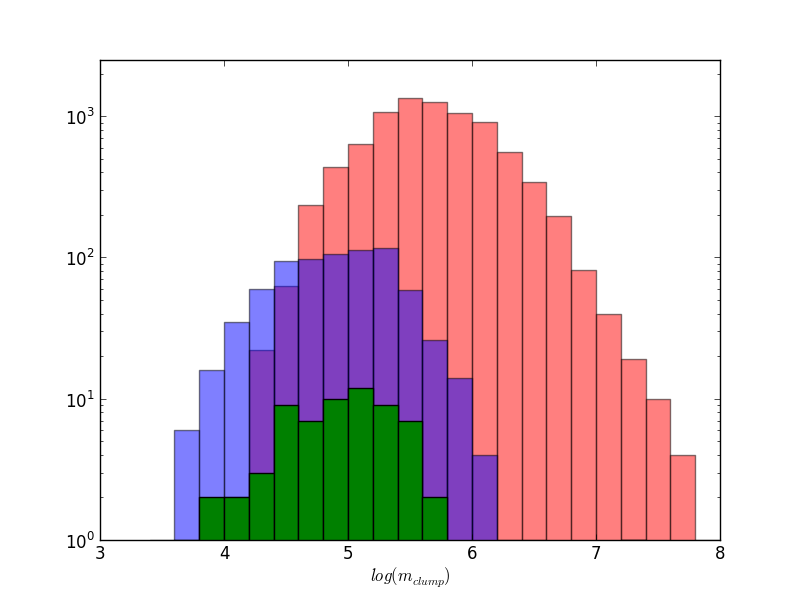}
\caption{The distribution of the mass of the clumps at different times from $t_1$ to $t_3$. The unit of the clump mass $m_{clump}$ is solar mass M$_{\odot}$.
\label{fig:mass}}
\end{center}
\end{figure}

At $t_2$, the typical size of the clumps is between 20 and 100 pc. The mass of the clumps ranges from thousands to millions of M$_{\odot}$, with a distribution shown in Figure~\ref{fig:mass}. Also shown are the clump mass distributions at $t_1$ and $t_3$.  The total number and mass of the clumps grows with time. The initial growth of the total number of clumps is exponential ($\sim e^{0.18t}$ from $t_1$ to $t_2$). The distribution is approximately log-normal; however, we note that the lower limit of the clump mass is numerical: given the typical clump density ($\sim \, 10^{-22}$ g cm$^{-3}$), and the minimum clump size in our selection criteria (6 times the minimum cell volume in our simulation), the minimum cell mass is a few thousand M$_{\odot}$. This effect can be seen in Figure~\ref{fig:mass} when comparing the lower end of the mass distribution at different times. Since we decrease our resolution by a factor of 2 between $t_2$ and $t_3$, the numerical limit of the smallest clump size increases, resulting in a slightly higher cut-off at the lower end of the mass distribution at $t_3$ compared to the previous times.

\begin{figure}
\begin{center}
\includegraphics[scale=.81]{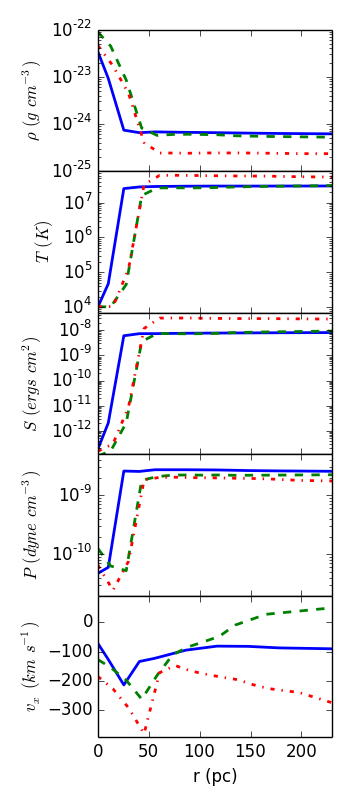}
\caption{The gas properties (from top to bottom: density, temperature, entropy, pressure and x-velocity) along a line from the center of the clump outwards in the positive x direction for three randomly selected clumps in three columns. The masses of the three clumps shown in solid blue, dashed green and dash-dotted red lines are: $5.8$, $9.5$ and $14.4$ $\times 10^5$ M$_{\odot}$ respectively. The structures are quite similar despite the difference in the mass of the three clumps.
\label{fig:3by5}}
\end{center}
\end{figure}

We also briefly examine the structure of the clumps, although we note their internal structure is -- at best -- marginally resolved.  For a representative sample of clumps, we plot the gas properties along a line drawn from the center-of-mass outwards. We choose the direction of the line to be along the positive x axis for simplicity, but this should be considered an arbitrary direction from the clump center since the clumps have no significant systematic motion along the x direction. Since the clumps are not perfectly spherically symmetric, the centers are not always well defined, but most of the clumps show the same pattern seen in Figure~\ref{fig:3by5}, where we present three individual clumps as examples and plot the gas properties along the vector we described earlier. 

The ICM around the clumps is often rather smooth. Immediately surrounding the clumps, there is a very thin transition layer where the density shows a sharp increase and temperature and entropy (and to a lesser extent, pressure) drop drastically.  Perhaps surprisingly, the clumps are not in pressure equilibrium, but systematically have lower pressures then their surroundings.  This pressure gradient drives an inflow; however, due to the rapid cooling, pressure equilibrium cannot be established and the clump mass grows with time.

The inflow velocity reaches its maximum ($v_x$ is most negative) where the pressure gradient is the steepest. Inside the clumps, the temperature decreases until it reaches $T_{\rm floor}\sim 10^4$ K; the pressure often increases slightly towards the center of the clump due to the accumulation of material; the inflow velocity decreases (and would change sign if we were to show the other side of the plot along the negative x direction). The central velocity of the clump is not subtracted from $v_x$ due to the difficulty of defining the clump center.  We note again that the internal structure of the clumps in these simulations must be viewed with much caution, both because they are poorly resolved, and also because of missing physics (magnetic fields and heat conduction, in particular, are likely to effect this structure).

\subsection{Comparison with Observations}\label{sec:observations}

\begin{figure*}
\begin{center}
\includegraphics[scale=.43,trim=1cm 0cm 0cm 0cm, clip=true]{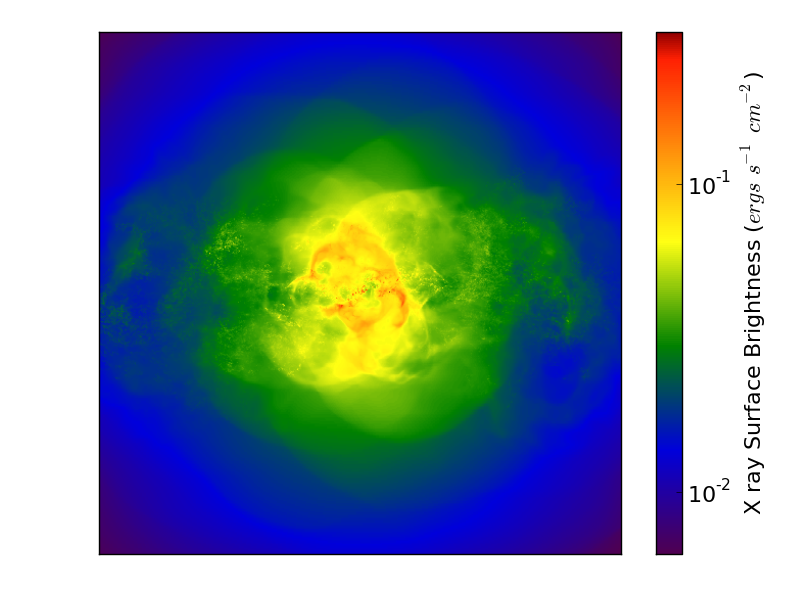}
\includegraphics[scale=.43,trim=2cm 0cm 0cm 0cm, clip=true]{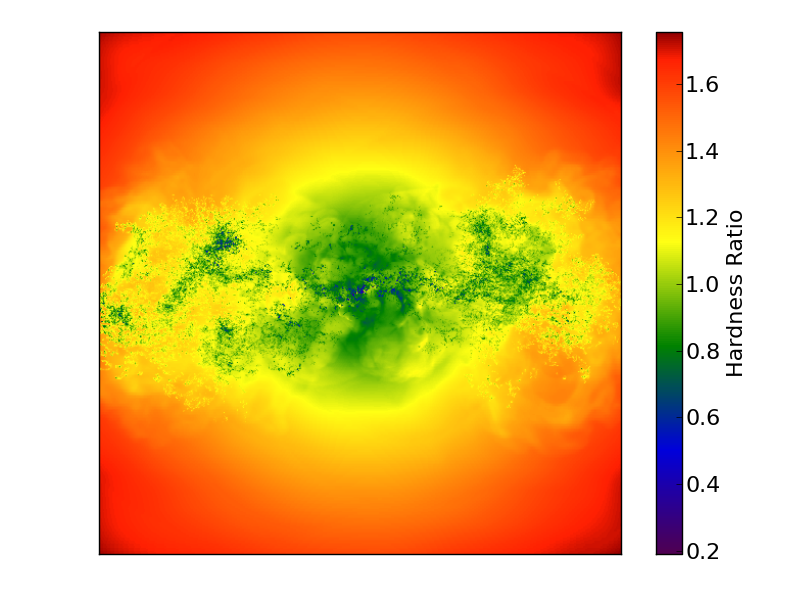}
\includegraphics[scale=.43,trim=1cm 0cm 0cm 0cm, clip=true]{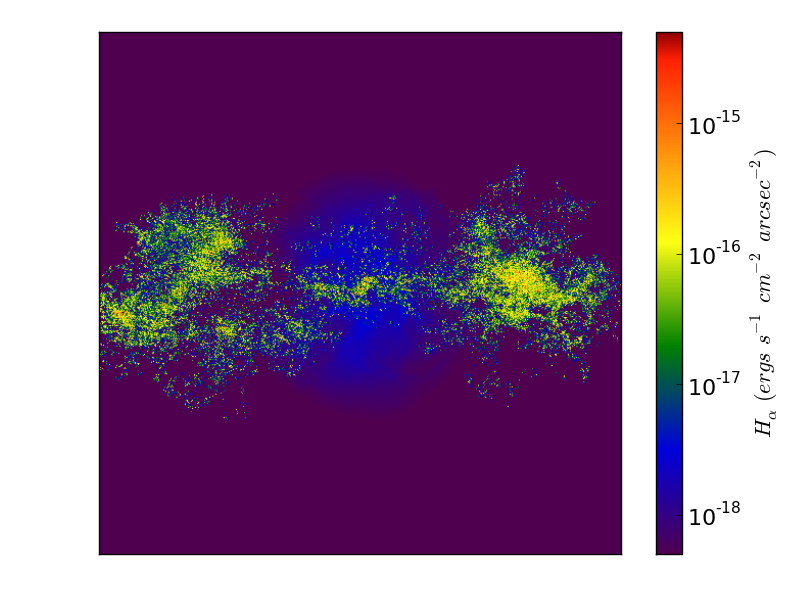}
\includegraphics[scale=.43,trim=2cm 0cm 0cm 0cm, clip=true]{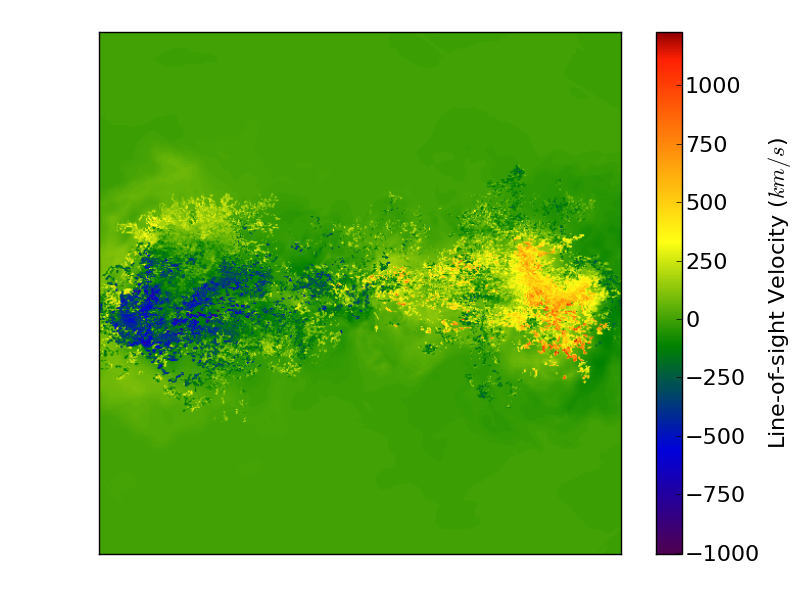}
\caption{The simulated observations of the central 40 kpc area at $t_3 = 410$ Myr. The maps shown here are: 0.5-3 keV X-ray flux (top left), X-ray hardness ratio (1.5-7.5 keV map divided by 0.3-1.5 keV, top right), $H{\alpha}$ emission (bottom left) and $H{\alpha}$ emissivity weighted average line-of-sight velocity of the gas (bottom right). The observer's angle is about $52^\circ$ from the z axis in the y-z plane. \label{fig:Halpha}}
\end{center}
\end{figure*}

In order to compare our simulation results to the observations more directly, we generate simulated observational maps of the X-ray surface brightness, the X-ray hardness ratio (1.5-7.5 keV to 0.3-1.5 keV), $H{\alpha}$ surface brightness and the $H{\alpha}$-weighted gas velocity map in the central 40 kpc box. The X-ray and $H{\alpha}$ emissivities are computed through interpolation of a multidimensional table generated with version 13.00 of Cloudy, last described by \citet{Cloudy}, assuming coronal equilibrium and no photo ionization, which means that our $H{\alpha}$ brightness should be considered a lower limit. Since our cooling function is truncated at $T_{floor}$ very close to $10^4$ K, a lot of cold gas sits at that temperature which happens to be the peak of the $H{\alpha}$ emissivity. To avoid the artificial enhancement of the $H{\alpha}$ luminosity due to the accumulation of the gas near $T_{floor}$, we re-run the simulation for a short period of time with a new cooling curve that extends to 300 K before we generate the $H{\alpha}$ map. The observational angle should be arbitrary with respect to the jet propagation direction. 

We show in Figure~\ref{fig:Halpha} an example of the simulated observations at an angle of $52^{\circ}$ from the z axis in the y-z plane at $t_3$, a time corresponding to the last panel in Figure~\ref{fig:projection_density} and Figure~\ref{fig:projection}. The clumps show up clearly on the $H{\alpha}$ map (bottom-left panel), but are largely absent in the X-ray map (top-left panel).  In particular, the largest groups of cold clumps, at a projected distance of about 15 kpc shows only very low surface brightness features in the X-ray image.

However, the H$\alpha$ maps do correlate spatially with areas of low hardness ratio in the X-ray hardness ratio map (top-right panel). This link between the optical nebulae and the soft X-ray filaments has been observed in nearby cool-core clusters \citep[e.g.,][]{Fabian2003, Sparks2004, Crawford2005}.  Although this feature is often argued to be a product of heat conduction, our simulations do not include heat conduction and yet still find this correlation. An explanation for this is that the soft X-rays comes from the gas in the transition layer surrounding the cold clumps (see Figure~\ref{fig:3by5}) and the gas that is stripped out of the cold clumps by the jets before it gets mixed into the hot ICM.

The X-ray map largely shows features related to the shocks. There is also an enhanced X-ray emission in the central 5 kpc region, coming from the low entropy gas in the center of the cluster. This region also shows a low X-ray hardness ratio on the hardness map, also consistent with observations \citep[e.g.,][]{Gitti2011}. 

The total $H{\alpha}$ luminosity of the cluster core at $t_3$ is about $6.4 \times 10^{41}$ ergs s$^{-1}$ and the total amount of cold gas is about $1.8 \times 10^{10}$ M$_{\odot}$, in good agreement with the observed relationship between $H{\alpha}$ luminosity and the total mass of the cold gas \citep{Edge2001, Salome2003}. 

The $H{\alpha}$ emissivity weighted average line-of-sight velocities of the cold clouds are shown in the last panel of Figure~\ref{fig:Halpha}. The typical velocity magnitudes range from 0 to a few hundred km/s, and large spatial variations are seen on both sides of the SMBH, i.e. positive and negative velocities are found next to each other. These features are consistent with the velocity field observed in $H{\alpha}$ and CO in Perseus Cluster \citep{Conselice2001, Salome2006}. \citet{Sharma12} has also found some newly formed cold clumps with non-zero velocity in their simulations. There is some systematic motion of the cold gas in our simulations with the left side being more negative and the right more positive. This is because, at the early stage of clump formation, many of the newly formed clumps are still moving outwards along the jet direction. We are seeing the projected systematic motion in Figure~\ref{fig:Halpha}. Recent ALMA observations have revealed systems with $\sim 10^{10} \; \rm M_{\odot}$ molecular gas flowing out of the BCG at velocities of hundreds of km/s accompanying AGN activities in the center of cool-core clusters \citep{ALMA1, ALMA2}, similar to what we find in our simulations around $t_2$ - $t_3$. However, we do not expect all clusters to show this feature as this phase only lasts for a few hundred Myr and we will discuss the later evolution in Section~\ref{sec:later}.

\citet{McDonald10} found that the largest radius at which $H{\alpha}$ emission is detected in a cluster ($R_{H{\alpha}}$) never exceeds its cooling radius $R_{\rm cool}$, defined as the radius at which the gas cooling time is 5 Gyr. In the Perseus cluster, as well as in our simulation, $R_{\rm cool} \sim$ 80 kpc. The clumps are never seen at radii larger than $R_{\rm cool}$ at any time in our simulations. This ``hard limit'' of $R_{H{\alpha}}$ in our simulations is a natural result of the combination of two factors: the slowing down of the jets and the larger $t_{TI}/t_{ff}$ at larger radii.

\section{Discussion}\label{sec:discussion}

\subsection{Later Evolution}\label{sec:later}

We have discussed the burst of formation of the cold clumps in Section~\ref{sec:formation}, which applies to the early stages of our simulation, from $t_1$ to $t_2$. While new clumps are forming, the lower pressure of the clumps that have already formed drives an inflow as we have shown in Figure~\ref{fig:3by5}, allowing ambient hot gas to cool onto them, increasing their mass (Figure~\ref{fig:mass}). 

As shown in Figure~\ref{fig:projection}, the regions containing clumps are automatically refined to the highest level due to their high density. Therefore, the existence of a large amount of clumps slows down the simulation dramatically. In order to see the later evolution of the clumps, we lower the resolution by setting the maximum refinement level to $l_{\rm max} = 10$ after $t\sim 425$ Myr, effectively changing the highest resolution in the simulation to 30 pc. This change in resolution suppresses formation of the lowest mass clumps, but we found it otherwise to have only a mild change on the evolution of the amount of cold gas.

The clumps, once formed, lose pressure support and effectively decouple from the hot ICM, moving almost ballistically. Some of them still move outwards for a while due to their initial positive radial velocity, but shortly they rain back down towards the center of the cluster. They reach the center of the cluster within roughly a free-fall time and then oscillate or rotate around the center while being accreted to the SMBH. New clumps still keep forming at radii of a few to a few tens of kpc while more cold gas accumulates in the center. Meanwhile, the jet power goes up due to the accretion of these cold clumps, heating up the core of the cluster and elevating $t_{TI}/t_{ff}$. New clump formation becomes more and more rare and eventually the ICM becomes stable in spite of the jet perturbation. After a few Gyr, most of the cold gas has settled to a rotating disk of a few kpc in radius around the SMBH and clumps are rarely seen at larger radii.

This result is consistent with what \citet{Cattaneo2007} and \citet{Gaspari2012} have found in their simulations. We will discuss the later evolution in more detail in Paper III. Observationally, many nearby cool-core clusters have a rotating $H{\alpha}$ structure in the central $< 10$ kpc region \citep{McDonald12}. \citet{McDonald10, McDonald11} found that the clusters with extended $H{\alpha}$ filaments have lower core temperature and entropy than those that do not have $H{\alpha}$ detection or have only nuclear $H{\alpha}$ emission. This is consistent with what we see in our simulations: at the beginning when the temperature is still high, there are no clumps; the clumps form when the temperature is the lowest; AGN feedback heats up the core and cold gas only exists in the central region at late times. It is not clear whether CC clusters all experience this sequence or they can experience multiple bursts of clump formation that are not seen in our simulations due to the lack of some physics such as merger events, magnetic fields or heat conduction.

On the other hand, the lack of filaments outside the central $< 10$ kpc area at late times in our simulations shows that jets cannot lift up the coolest gas from the bottom of the potential to larger radii.

\subsection{Simulations with Different Parameters and Resolutions}

We observe the formation of cold clumps in all of our simulations with different parameters and feedback mechanisms and therefore find clump formation to be a robust outcome of AGN feedback in cool-core clusters.  In this section, we remark on some of the differences in these simulations.

In the simulation with $f=1$, i.e. pure kinetic feedback, the clumps initially form at slightly larger radii, $\sim10-20$ kpc, and quickly extend to $20-40$ kpc. This is likely due to the faster velocity of the jets in this simulation, where all the energy comes out as kinetic power, while in our standard run half of it is in the form of thermal energy. The jets are able to perturb gas at a larger distance from the center and to also bring the low entropy gas further before it cools into clumps. The angular distribution of the clumps is similar to that in our standard run, with most of them formed along the jet propagation direction.

In our low resolution simulations with $N_{\rm root}=64$ and $l_{\rm max}=10$ -- which gives $\Delta x_{min}\approx 244$ pc, comparable to \citet{Gaspari2012} -- we find that the clumps/filaments are fewer and bigger due to the lower resolution, but also more isotropically distributed. This is because when we lower the resolution, we effectively increase the physical width of the jets, which can directly perturb a much larger area.  In addition, larger clumps are easier to disrupt.

A comparison low resolution run with an order of magnitude higher feedback efficiency ($\epsilon = 0.01$) gives similar results to our cannonical run. The total amount of cold gas is a few times lower. This occurs because, when large amount of cold gas form and fall to the center, the jet power increases more quickly and thus the core is heated up more quickly, reducing the additional formation and growth of the clumps.  We will discuss the long term evolution and heating/cooling balance in more detail in Paper III.

\subsection{Limitations of our Model}\label{sec:limits}

There are a number of important pieces of physics are not included in our simulations, including heat conduction, magnetic fields, viscosity, and star formation. As we have discussed in Paper I (and is now generally well-accepted), thermal conduction alone is unable to prevent the global cooling catastrophe, but it does increase the temperature at radii from a few to a few tens of kpc. It could also slow down the growth of the clumps after they are formed, or suppress them from forming in the first place.  Both of these would most likely reduce the amount of cold gas, although we have not explored this in these simulations. It may also help the jets to distribute the heat more evenly throughout the core.

Magnetic fields have been suggested as a mechanism to support the filaments against gravity and help maintain their thin long structure \citep{Fabian2003, Fabian2008, Lim2009}. We find that in our simulations, without magnetic fields, the shape of the cold gas clouds are more clumpy and less filamentary than observations indicate. The magnetic pressure in the filaments may slow down their growth after they are formed. On the other hand, magnetic fields would also suppress heat conduction across the field lines, making the filaments less vulnerable to destruction due to conduction. Therefore, it is not clear weather magnetic fields will increase or reduce the amount of cold gas. Magnetic fields will also affect the way heat conduction operates, giving rise to new instabilities that may impact these results.  We would like to explore the effect of magnetic fields in future work.

Signatures of star formation are observed in many cool-core clusters, which also seems to correlate with the presence of $H{\alpha}$ filaments \citep{Hicks2005, Hicks2010, McDonald11b}. Star formation can be a heating and ionizing source for the filaments locally \citep{McDonald12} while on a larger scale, it can drive turbulence in the cluster \citep{Revaz2008}. Again, we will leave this for future studies.

\section{Conclusion}\label{sec:conclusion}

We have carried out a set of high-resolution (15-30 pc minimum cell size) AMR simulations of a cool-core cluster including a model for AGN feedback.  We start the simulation from an initial configuration based on the Perseus cluster, employing AGN feedback in the form of precessing jets with an energy input driven by the amount of cold gas within 500 pc of the SMBH.  We find the following results.

(1) After about 300 Myr, cold clumps start to form along the jet propagation direction at a distance of 5-10 kpc from the cluster center. The ICM is still (albeit only marginally) linearly stable at the time of clump formation, but the jets perturb gas along its path in a non-linear way. By injecting tracer particles into the flow, we find that the gas that is about to cool into clumps has a radial velocity which is systematically larger than similar gas parcels which do not cool.  This indicates that low entropy (but still hot) gas is lifted up by the jets from a few kpc to larger distances, where the local free-fall timescale becomes comparable to the thermal instability timescale of the gas, allowing it to cool into dense clumps. In particlar, we stress that the gas does not cool spontaneously out of the hot ICM due to a linear thermal instability, nor can jets lift up signifcant amounts of cold gas from the center of the cluster. Instead the cold filaments form as a natural result of the interaction between the AGN jets and the cooling ICM in the core of the cluster.

(2) We confirm the work of \citet{Sharma12} that cold clumps can cool out of the flow when the ratio of the thermal instability time to the free-fall time ($t_{TI}/t_{ff}$) becomes sufficiently low; however, with a more realistic model for AGN feedback, we see a more nuanced picture.  In particular, we find that strong perturbations are required for this mechanism to work: this is demonstrated by the fact that clumps form only in the cone of gas directly affected by the jet, despite the fact that the ratio of $t_{TI}/t_{ff}$ is constant at fixed radius inside and outside of the jet cone.  There are indications that the exact value of $t_{TI}/t_{ff}$ required for the formation of cold clumps depends on the level of the turbulence, with stronger perturbations permitting cold condensation with higher ratios.  When the flow is smooth, clumps do not form at all (Paper I).

(3) The overall morphology, the spatial extension and the velocity field of the clumpy structure in our simulations generally agree with the observations. We also find a correlation between the simulated $H{\alpha}$ emission and the soft X-ray maps which is seen in nearby cool-core clusters. However, our clumps generally lack the filamentary nature seen in observations, probably because we are missing magnetic fields.

In a forthcoming paper (Paper III) we will explore the heating of the gas via this form of AGN feedback, investigating if it can lead to a realistic model in which heating balances cooling in a way consistent with observations.

\acknowledgments

We acknowledge financial support from NSF grants AST-0908390, AST-1008134, AST-1210890 and NASA grant NNX12AH41G,  as well as computational resources from NSF XSEDE and Columbia University.

\end{document}